\documentstyle[11pt,newpasp,twoside,epsfig]{article}
\def\chem#1#2{$\rm{}^{#2}\kern-0.8pt#1$}
\def\reac#1#2#3#4#5#6{$\rm\,{}^{#2}\kern-0.8pt{#1}\,({#3}\,,{#4})\,
{}^{#6}\kern-0.8pt{#5}\,$}
\def\gsimeq{\,\,\raise0.14em\hbox{$>$}\kern-0.76em\lower0.28em\hbox  
{$\sim$}\,\,}  
\def\lsimeq{\,\,\raise0.14em\hbox{$<$}\kern-0.76em\lower0.28em\hbox  
{$\sim$}\,\,}  
\def\beqy{\begin{eqnarray}}
\def\eeqy{\end{eqnarray}}
\def\bmlet{\begin{mathletters}}
\def\emlet{\end{mathletters}}

\def\edcomment#1{\iffalse\marginpar{\raggedright\sl#1\/}\else\relax\fi}
\reversemarginpar

\begin{document}
\title{Stellar yields in CNO from rotating stellar models}
 \author{G. Meynet, A. Maeder and R. Hirschi}
\affil{Geneva Observatory, CH--1290 Sauverny Switzerland}
 
\begin{abstract}
For $^{12}$C and $^{16}$O, rotating models predict 
in general enhanced yields.
At high metallicity, the carbon and oxygen yields from the very high mass stars, which go through a WR phase, are little affected by rotation.
For $^{14}$N, rotation allows the production of important amounts of primary nitrogen in {\it intermediate mass stars}
at very low metallicity. The process invoked for this production is different, 
from the classically accepted scenario {\it i.e.} 
the Hot Bottom Burning (HBB) in Asymptotic Giant Branch (AGB) stars. Rotating models also predict important productions of primary $^{13}$C, $^{17}$O and
$^{22}$Ne at very low metallicity.
\end{abstract}

Rotation affects all the outputs of the stellar models (Heger \& Langer 2000; Maeder \& Meynet 2000a)
and in particular the stellar yields. We shall not detail here the physical mechanisms driven by rotation.
These are described in many other papers to which the reader may refer (Zahn 1992; Maeder \& Zahn 1998; 
Heger \& Langer 2000; Maeder \& Meynet 2000a).
We want to illustrate here, by a few numerical examples, how the effects of rotation on mixing and mass loss may affect
the quantities of new CNO elements synthesized and ejected into the interstellar medium. For this we shall use recent grids of
stellar models computed for three different metallicities: $Z=0.00001$ (Meynet \& Maeder 2002b), $Z=0.004$
(Maeder \& Meynet 2001) and $Z=0.020$ (Meynet \& Maeder in preparation).
All these models were computed with the nuclear reaction rates of the NACRE compilation
(Angulo et al. 1999; see also the review by M. Arnould in this volume).

\begin{figure} 
\plottwo{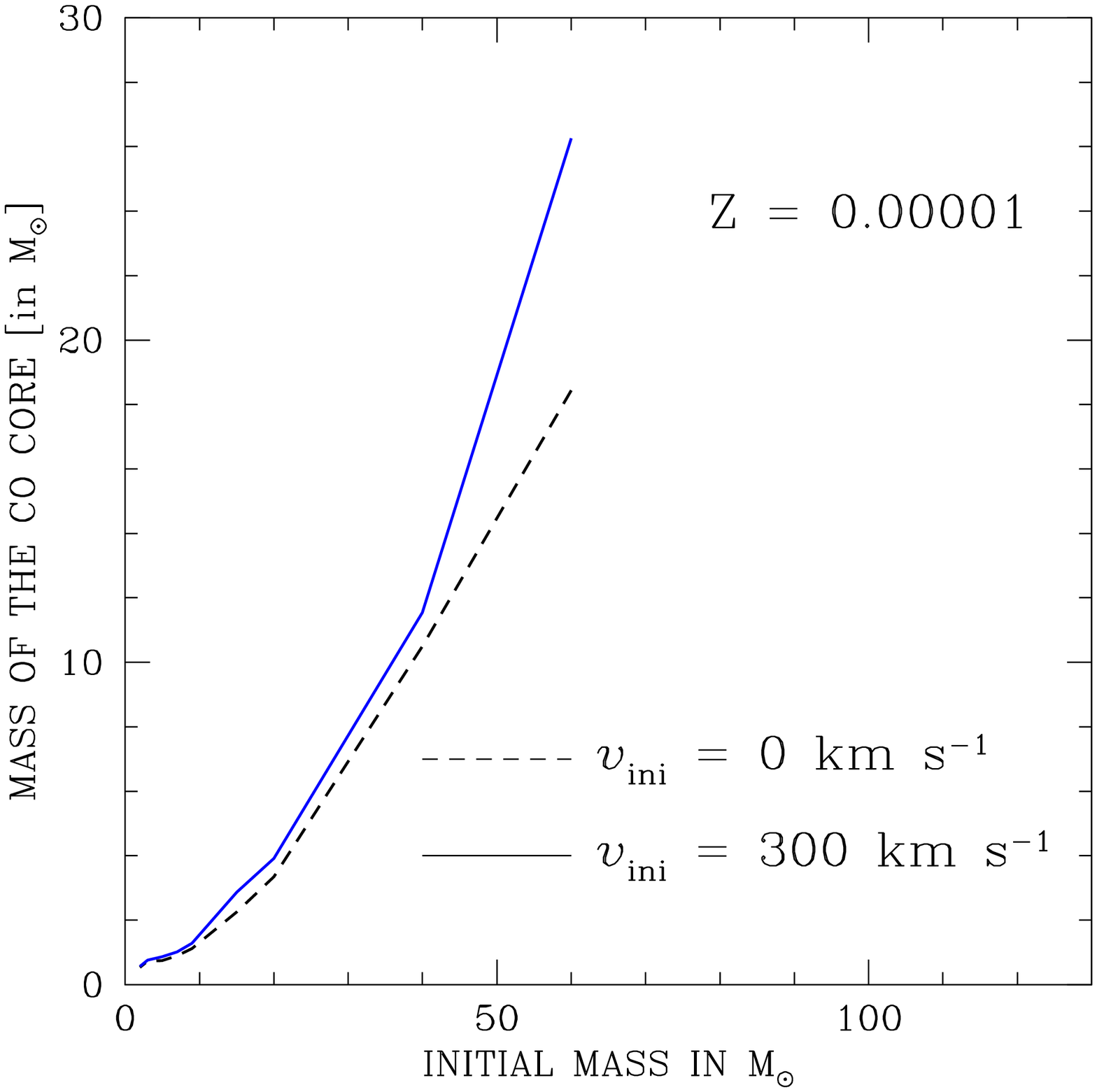}{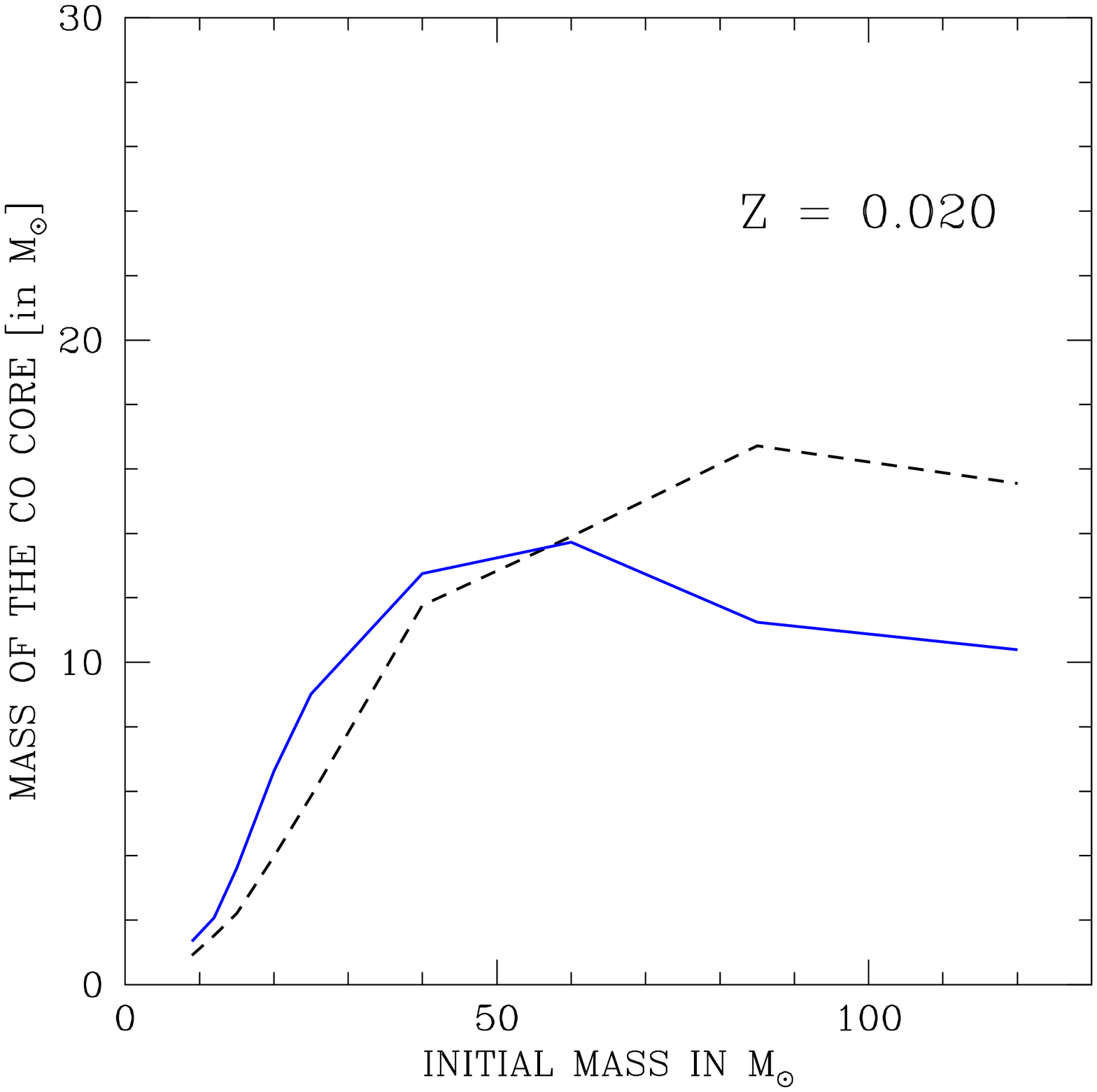}
\caption{{\it Left panel}: variation as a function of the initial mass of the mass of the carbon--oxygen (CO) 
cores obtained at the end of the C--burning phase. The CO--core is defined as the region interior 
to the shell where the mass fraction of carbon plus oxygen is superior to 0.75.
The models are those of Meynet and Maeder (2002b).
{\it Right panel}: Same as the left panel for solar metallicity models (models from Meynet \& Maeder, in preparation).}
\vspace{-0.5cm}
\end{figure}

\section{Rotation and the CO--core masses}

In Fig. 1, the variations of the CO--core masses as a function of the initial mass are indicated for various initial metallicities and for rotating and non--rotating stellar models. In stars with initial masses below 30--40 M$_\odot$, both at low and high metallicity, the CO cores are enhanced by rotation. As an example, at solar metallicity a rotating 20 M$_\odot$ stellar model with $v_{\rm ini}= 300$ km s$^{-1}$ has a CO--core mass equal to that of a non--rotating 26 M$_\odot$ stellar model. Thus one expects that rotating models will inject more carbon and oxygen than their non--rotating counterparts.

The increase of the CO--cores due to rotation is less marked at low metallicity.
Paradoxically this is a consequence of the more efficient rotational mixing at low $Z$ (Meynet \& Maeder 2002b). 
Indeed the more efficient mixing 
continuously replenishes in hydrogen the H--burning shell.
As a consequence the H--burning shell progresses less rapidly outwards and thus prevents the He--core mass and therefore the CO--core mass to increase as much as at
higher metallicity. 

For stars above 30--40 M$_\odot$, at solar metallicity, the CO--cores in rotating models are smaller
than in non--rotating ones. Due to rotational mixing, these
models enter into the Wolf--Rayet phase at an earlier stage of their evolution. During the Wolf--Rayet phase, which lasts longer in rotating models, the stars loose mass at a higher rate, producing smaller final masses. As we shall see below, the CNO yields for these high mass stars are little affected by rotation. 

\section{Rotation and mass loss by stellar winds}

\begin{figure} 
\plottwo{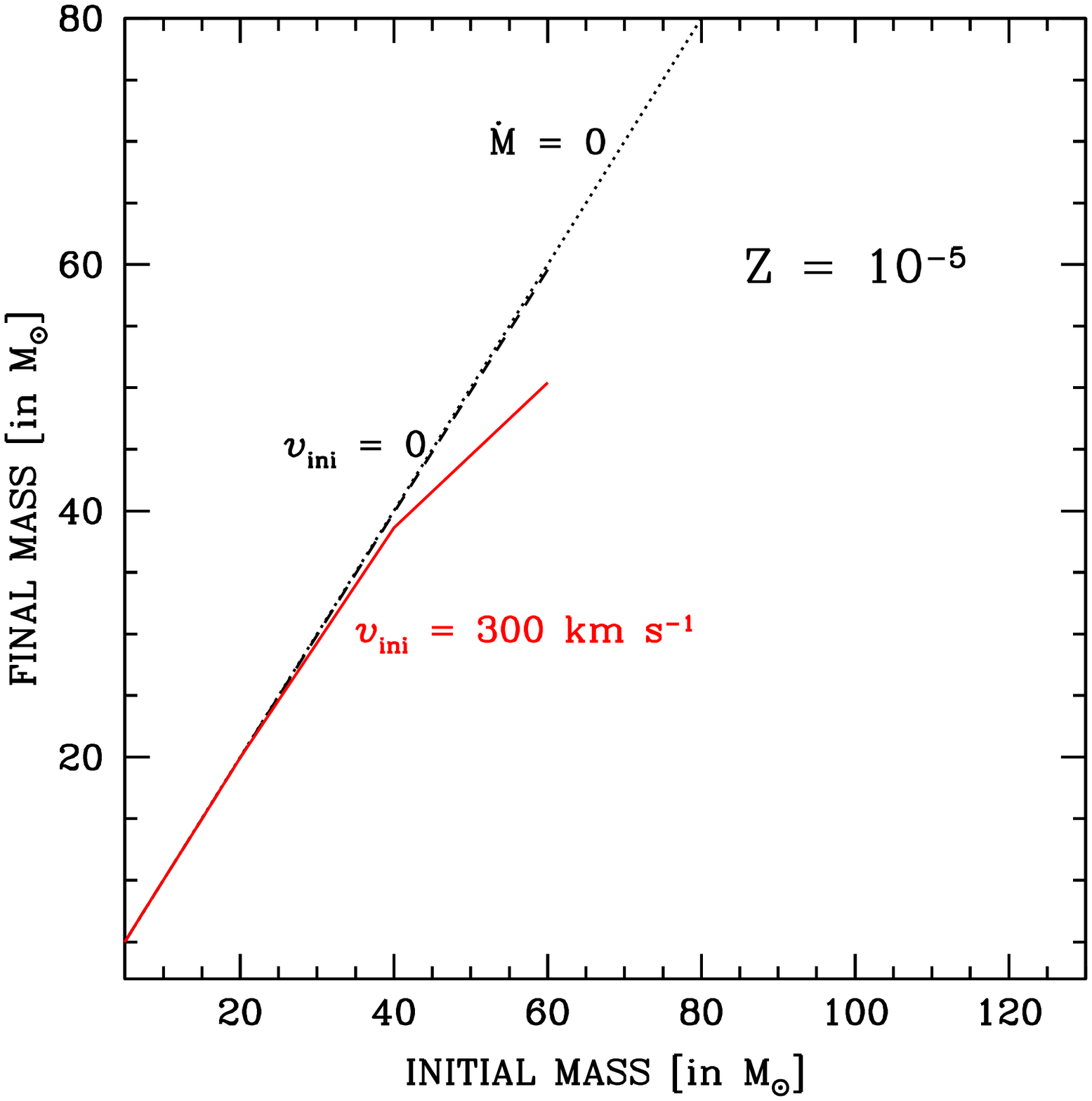}{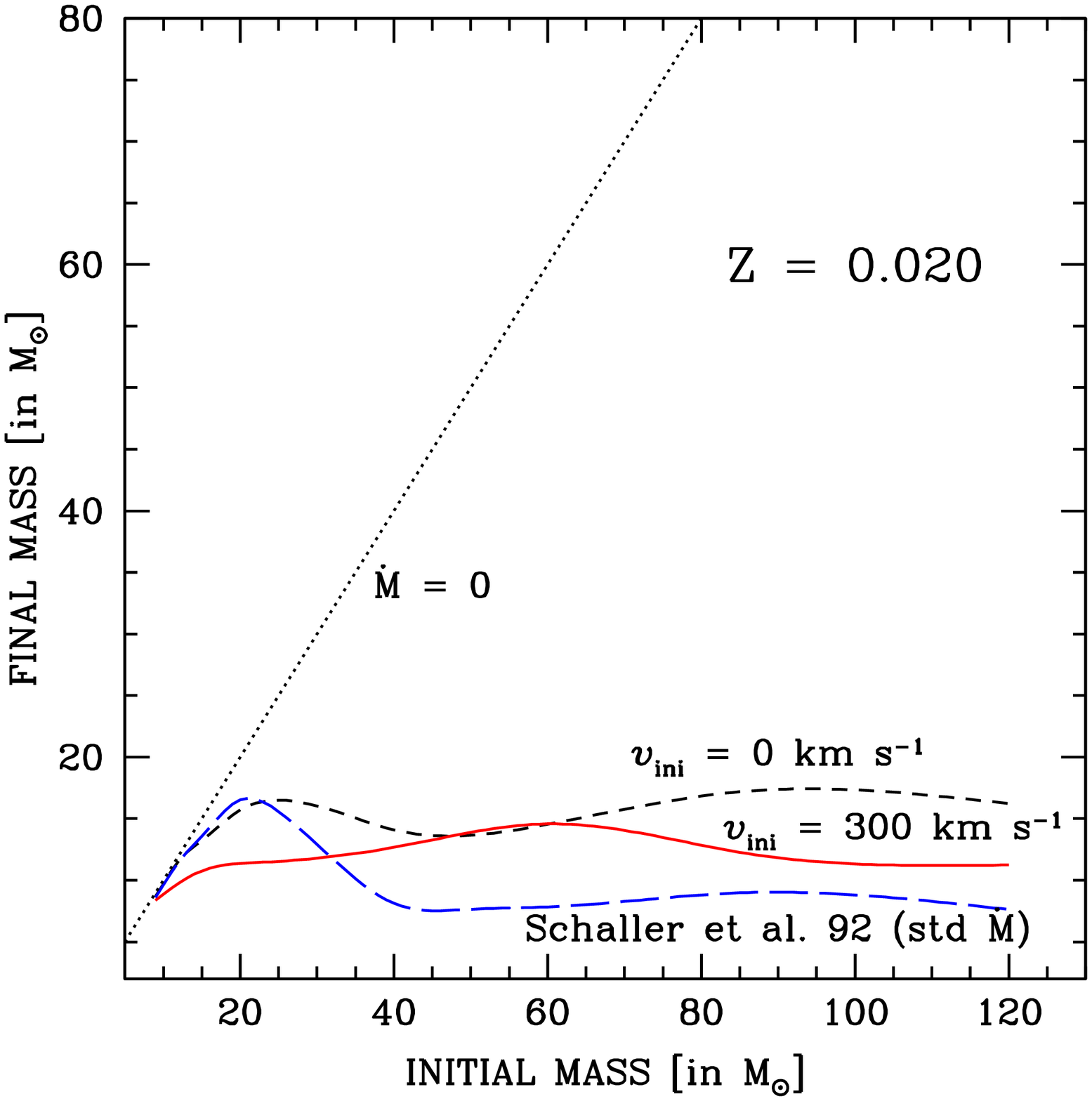}
\caption{{\it Left panel}: Relations between the final mass versus the initial mass at $Z=0.00001$. 
The cases with and without rotation are indicated. 
The dotted line with slope one would correspond to the case without mass loss. 
{\it Right panel}: Same as the left panel for models at $Z=0.020$, the relation obtained from
the models of Schaller et al (1992) is also shown.}
\end{figure}

In the framework of the radiative driven wind theory, it is possible to obtain an expression for the ratio of the mass loss rates from a rotating and a non--rotating star lying at the same position in the HR diagram (Maeder \& Meynet 2000b). Typically
for a 20 M$_\odot$ star model at the end of the MS phase, the enhancement factor due to rotation is at most 1.67 at break--up
while for a 60 M$_\odot$ stellar model, in the same conditions, the factor amounts to 3.78.  
Rotation does not only increase the quantities of mass lost by stellar winds but also induces anisotropies in the winds. Typically, polar winds are expected for fast rotating hot stars, which, as a consequence, will loose small amounts of angular momentum and thus
reach more easily the break--up limit (Maeder 2002).
Other effects induced by rotation may change the quantity of mass lost by
stellar winds:  the increase of the MS lifetimes, the bluer evolutionary tracks,
the easier Wolf--Rayet star formation.
  
Fig. 2 shows the final masses obtained for different initial masses, velocities and metallicities. 
At low metallicity, for stars with initial masses below 40 M$_\odot$ and with
$\upsilon_{\rm ini}$= 300 km s$^{-1}$, 
rotating and
non--rotating stellar models end with  final masses close to their initial values. Only in the case of the 60 M$_\odot$ there is a significant difference. One expects that the difference will increase when the mass increases and/or when the initial velocity increases. Concerning this last point it is worthwhile to recall here that
there are some indications that the relative number of fast rotators increases when the metallicity decreases (Maeder et al. 1999). This fact, together with the effect of rotation on both mixing and mass loss, may have interesting consequences
for the nucleosynthesis expected from the first stellar generations in the Universe.

At solar metallicity, mass loss has an important impact on the values of the final masses. All stars with initial masses 
above 40 M$_\odot$ end with final masses between 10--15 M$_\odot$. In general, as expected, rotating models produce smaller
final masses. One notes also that the new mass loss rates used in the recent computations
are smaller than those used
in the stellar grids of Schaller et al. (1992). This has some impact on the
carbon yields from the most massive stars as we shall see in Sect. 4.

\section{Effects of rotation on the carbon and oxygen yields}

Stellar yields at the metallicities $Z$ = 0.00001 and 0.004, for $^{4}$He, $^{12}$C, $^{14}$N and $^{16}$O  from rotating and non--rotating models,
in the mass range between 2 and 60 M$_\odot$, are discussed in Meynet \& Maeder (2002b). Let us recall here the main results. Figure 3 shows the yields in $^{12}$C for non--rotating and rotating stellar models ($v_{\rm ini}$ = 300 km s$^{-1}$) for two metallicities and for mass ranges between 2 and 120 M$_\odot$. 
For the metallicities considered here, the effects of the stellar winds are small.
We see that 
the yields in carbon are generally enhanced in the rotating models. The enhancement factors are equal to 1.3 and 2.4 for the 20 M$_\odot$ models
at $Z$ = 10$^{-5}$ and 0.004 respectively. 
Larger carbon yields in rotating models are a consequence of the greater CO--core masses in rotating models (see Sect. 1 above). Interestingly, the yields
from the rotating models are not very different from those computed by Maeder (1992) for $Z$ = 0.001 computed without rotation but with a
moderate overshooting, while the rotating models were computed without overshooting. Rotation, by enlarging the CO core masses, acts in this respect as an overshoot.

\begin{figure} 
\plottwo{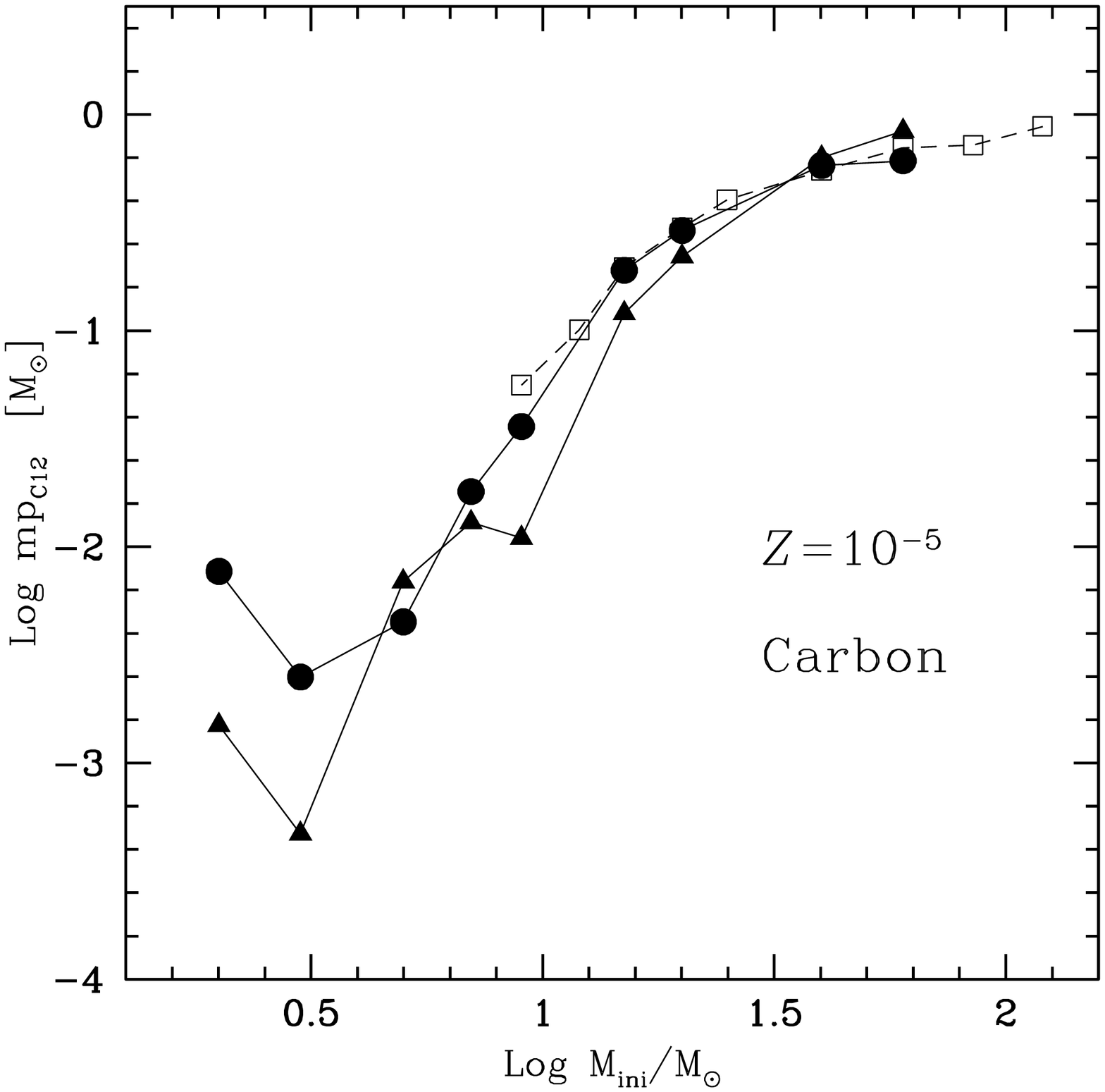}{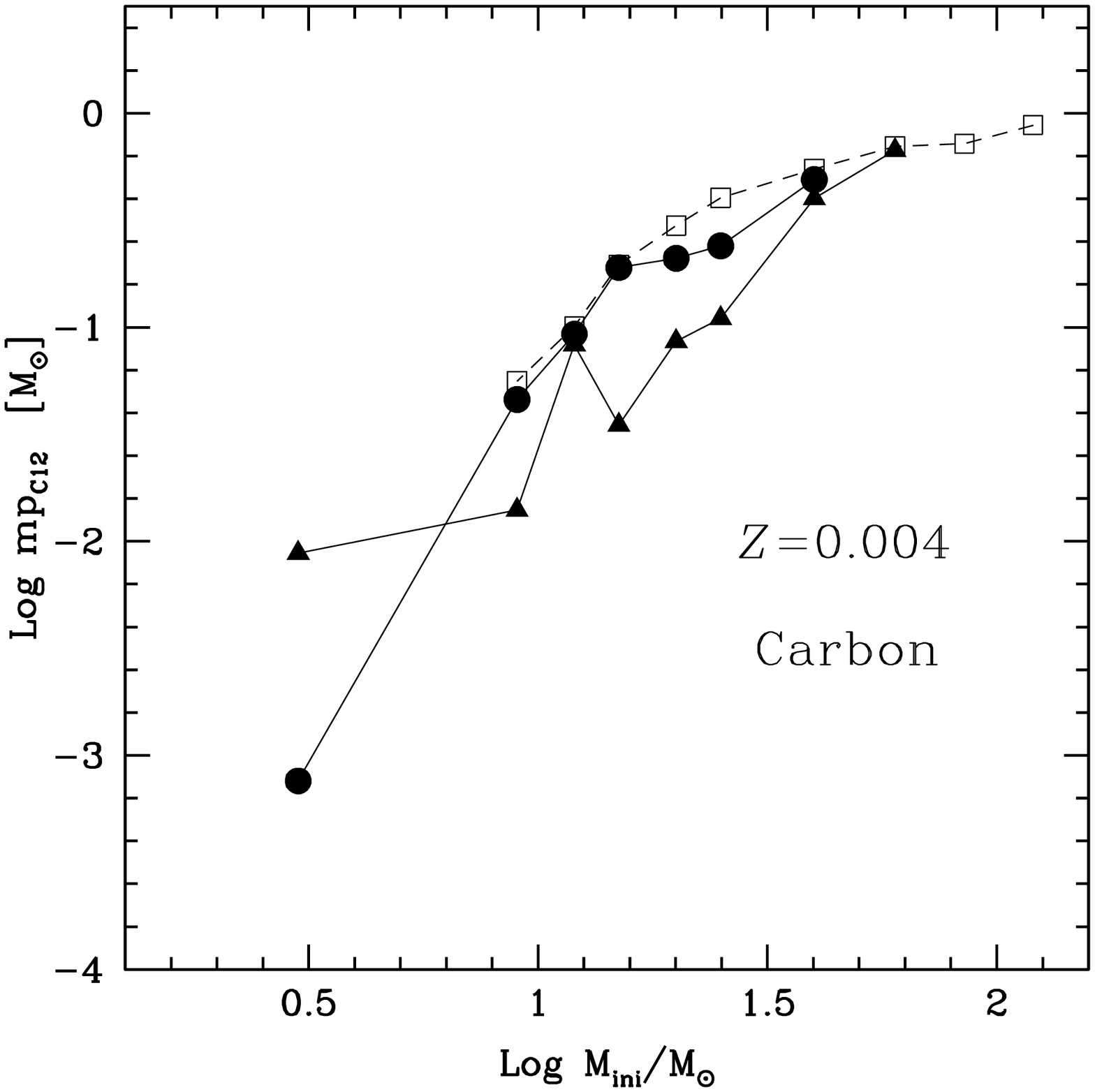}
\caption{{\it Left panel}: Yields of $^{12}$C for different initial mass stellar models: {\it black triangles},
non--rotating stellar models at $Z$ = 0.00001; {\it black circles}, rotating stellar models at $Z$ = 0.00001;
{\it empty squares}, non--rotating stellar models at $Z$ = 0.001 from Maeder (1992);{\it Right panel}: Same as the left panel for the metallicity $Z$ = 0.004.}
\end{figure}

\begin{figure} 
\plottwo{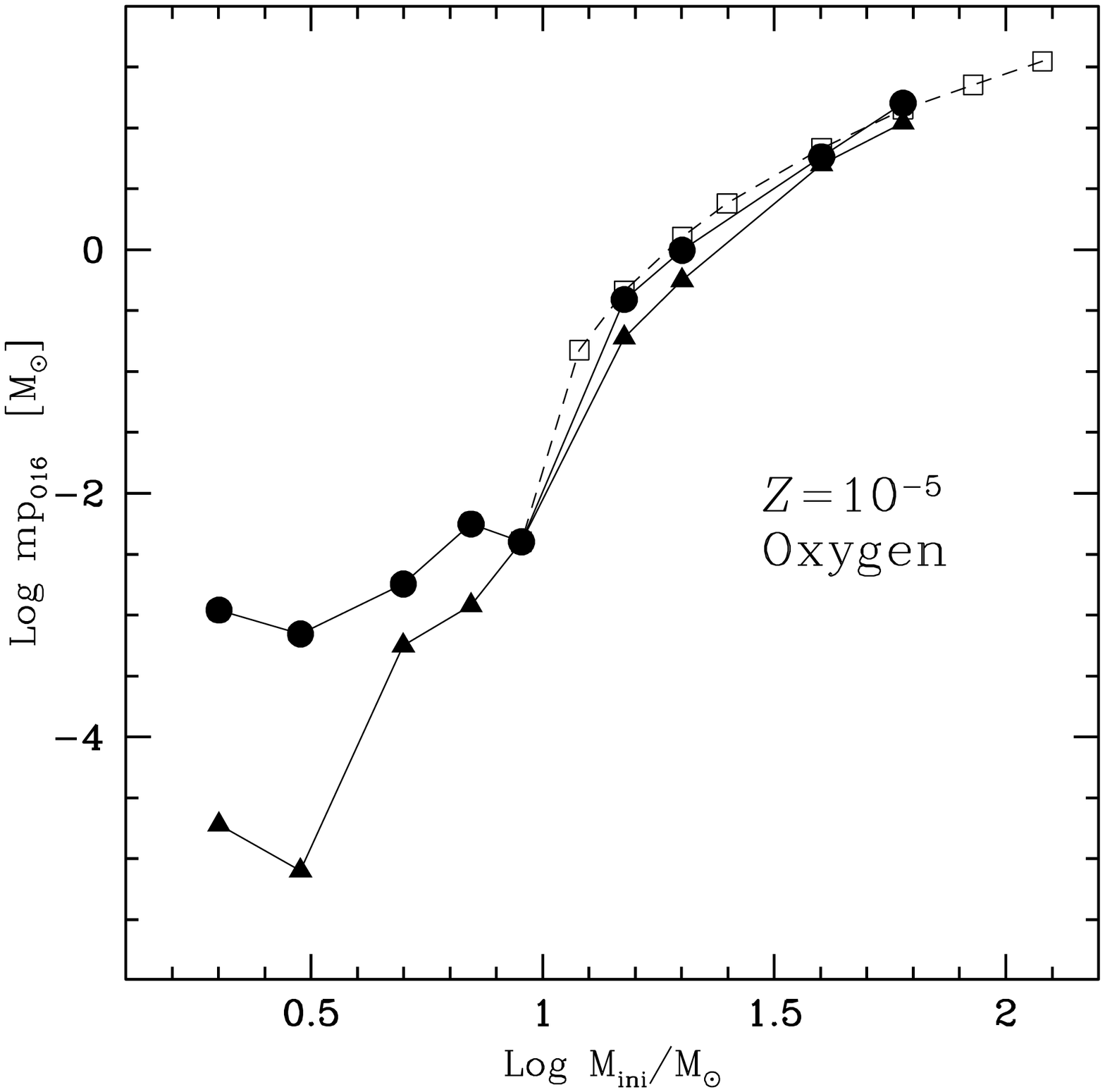}{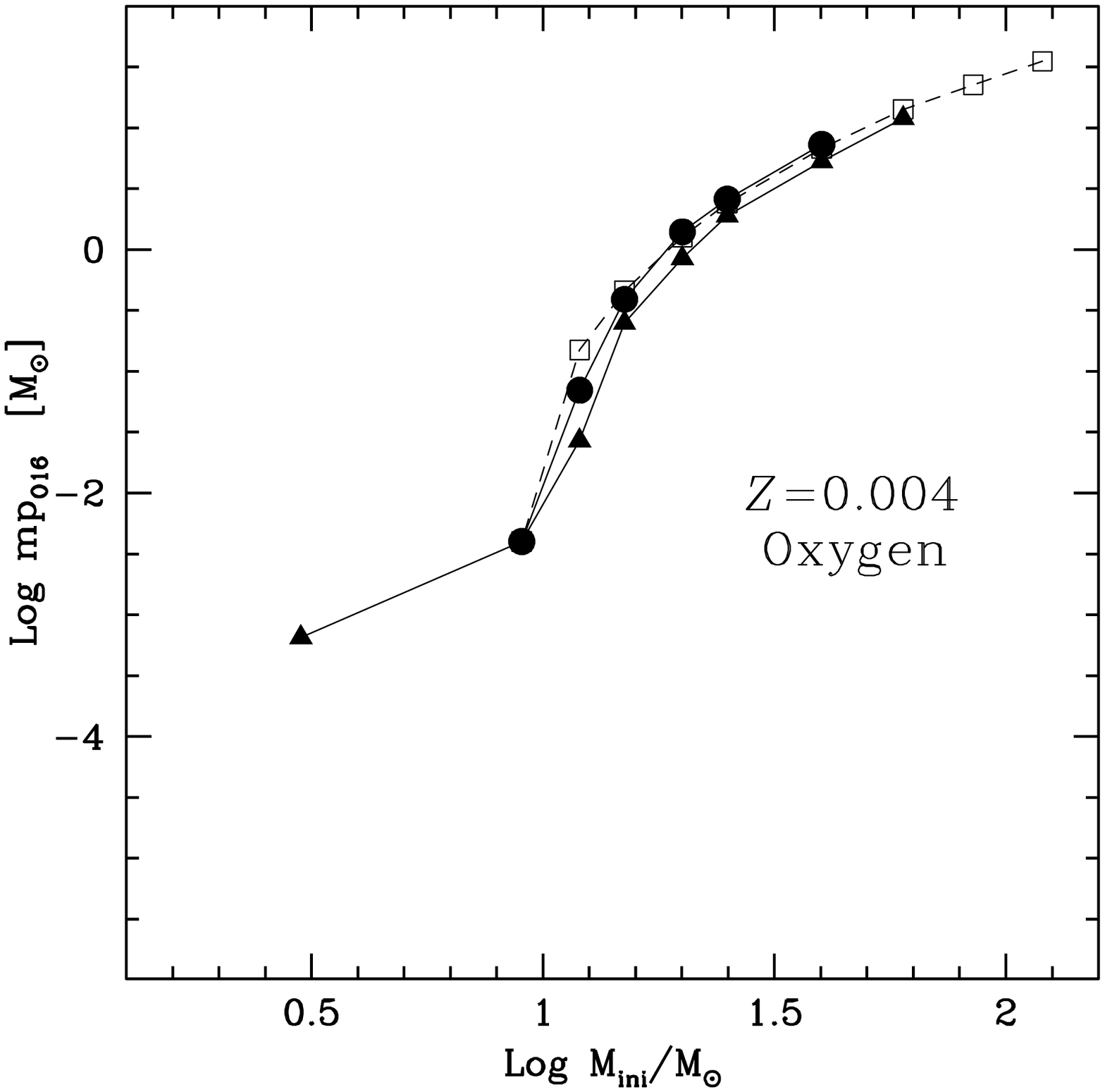}
\caption{{\it Left panel}: Yields of $^{16}$O for different initial mass stellar models: {\it black triangles},
non--rotating stellar models at $Z$ = 0.00001; {\it black circles}, rotating stellar models at $Z$ = 0.00001;
{\it empty squares}, non--rotating stellar models at $Z$ = 0.001 from Maeder (1992). {\it Right panel}: Same as the left panel for the metallicity $Z$ = 0.004.}
\end{figure}

Figure 4 shows the yields in $^{16}$O. As for carbon, the yields in oxygen are enhanced by rotation. For the 20 M$_\odot$, the enhancement factor is 1.8 at
$Z$ = 10$^{-5}$ and 1.6 at $Z$ = 0.004.
Again the yields from rotating models
(without overshooting) are very similar to those obtained by Maeder (1992).
  
At solar metallicity, the main difference with respect to the situation found at low metallicity essentially concerns the high mass star range
(stars with initial masses above about 40 M$_\odot$). For this mass range, mass loss by stellar winds becomes the dominant effect.
As a consequence the yields in carbon and oxygen in these stars  are much less affected by rotation. As a numerical example, the yields in carbon
and oxygen of a 60 M$_\odot$ star model differ by less than 11\% between the rotating and the non--rotating cases. 

It was proposed by Maeder (1992) that massive stars (going through a WC phase) may be important contributors of carbon
at high metallicity, an idea further sustained by various chemical evolution models (Prantzos et al. 1994; Carigi 2000;
Gustafsson et al. 1999). Since the mass loss rates have been reduced since the work of Maeder (1992), 
the present yields of carbon from WC stars are smaller, and those of oxygen greater. 
However these stars remain important sources of carbon at solar and higher metallicities.

\section{Effects of rotation on the nitrogen yields}

\begin{figure} 
\plottwo{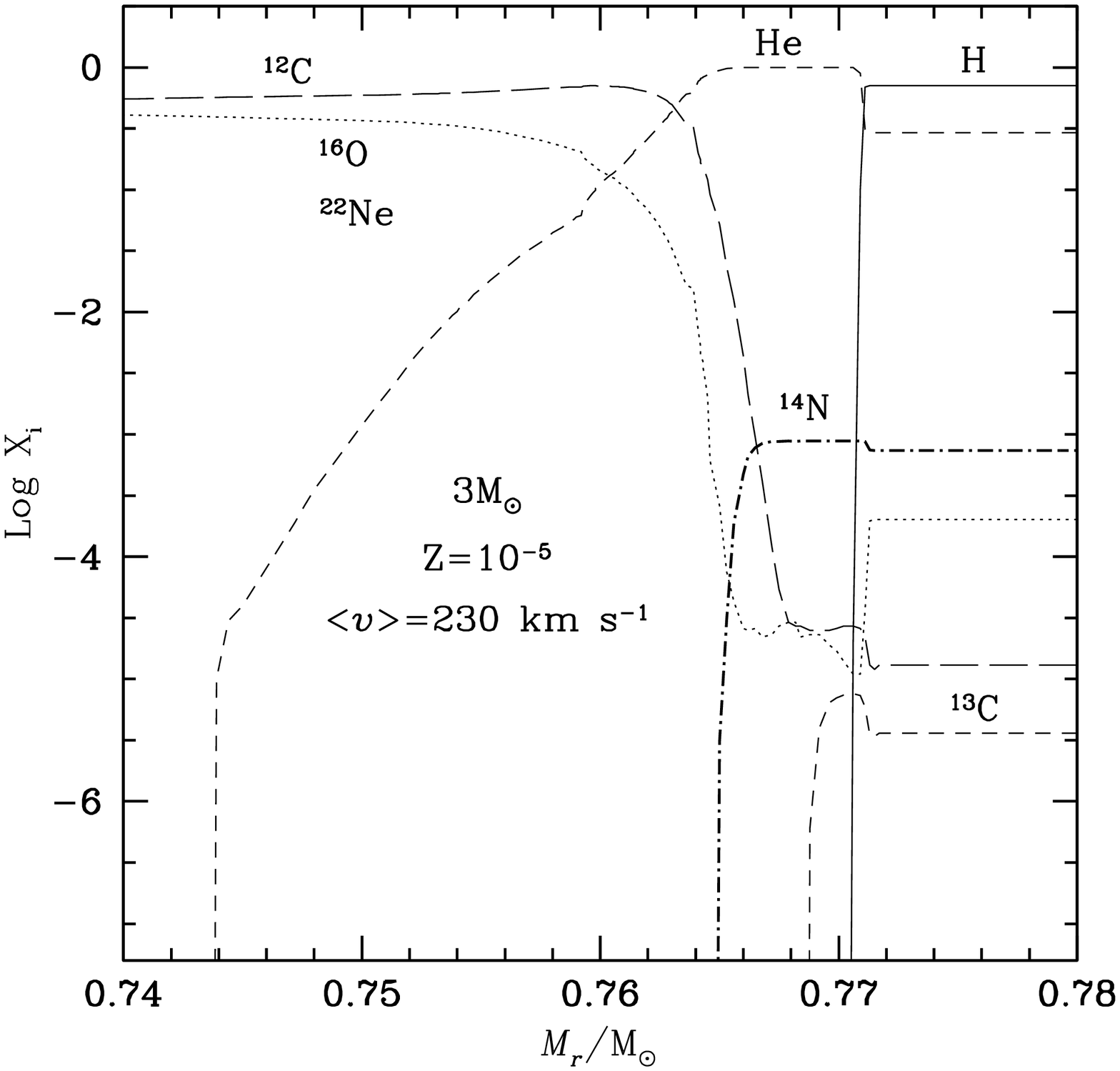}{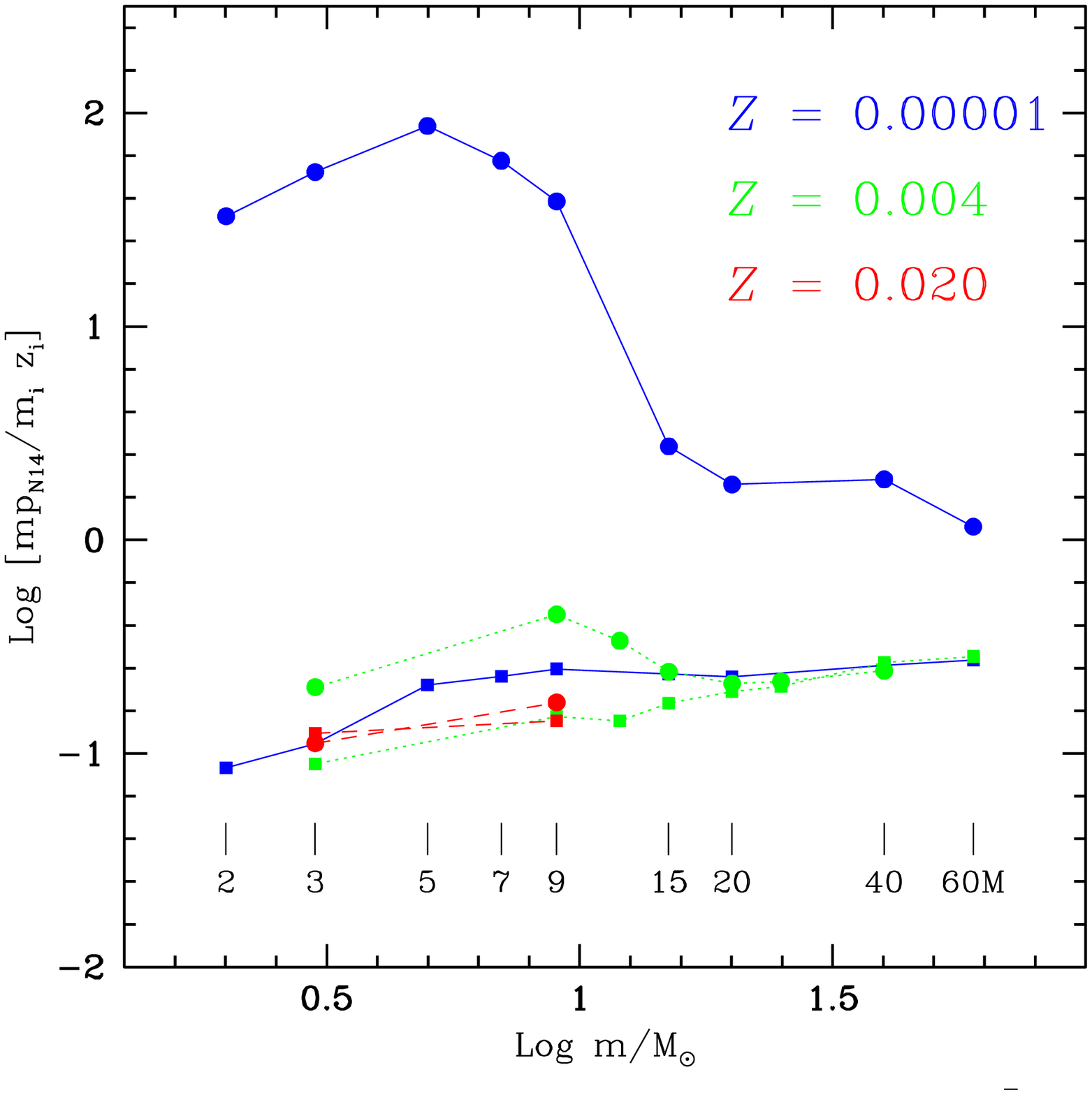}
\caption{{\it Left panel}: Chemical composition inside a rotating 3 M$_\odot$ star model at the beginning
of the TP-AGB phase. {\it Right panel}: Yields of $^{14}$N normalized to the initial metal content for different initial mass stellar models at various
metallicities. Black circles are for models with rotation, black squares are for non--rotating models, the continuous lines show
the models at $Z$ = 10$^{-5}$, the dotted lines, those at $Z$ = 0.004 and the dashed lines, those at $Z$ = 0.020.}
\end{figure}

Nitrogen is produced by transformation of carbon and oxygen through CNO cycle in H--burning zones.
Nitrogen is said to have a primary origin when the carbon (oxygen) used for its synthesis is produced by the star itself. 
The secondary channel corresponds to the case when
the carbon transformed into nitrogen is the one initially present in the star.
When nitrogen is primary, its abundance in the interstellar medium evolves in lockstep with other primary elements like, for instance, oxygen. In that case, the N/O ratio remains constant when the metallicity increases. When nitrogen is mainly produced by the secondary channel, the N/O ratio is expected to increase steeply with the increasing metallicity.

\begin{figure} 
\plottwo{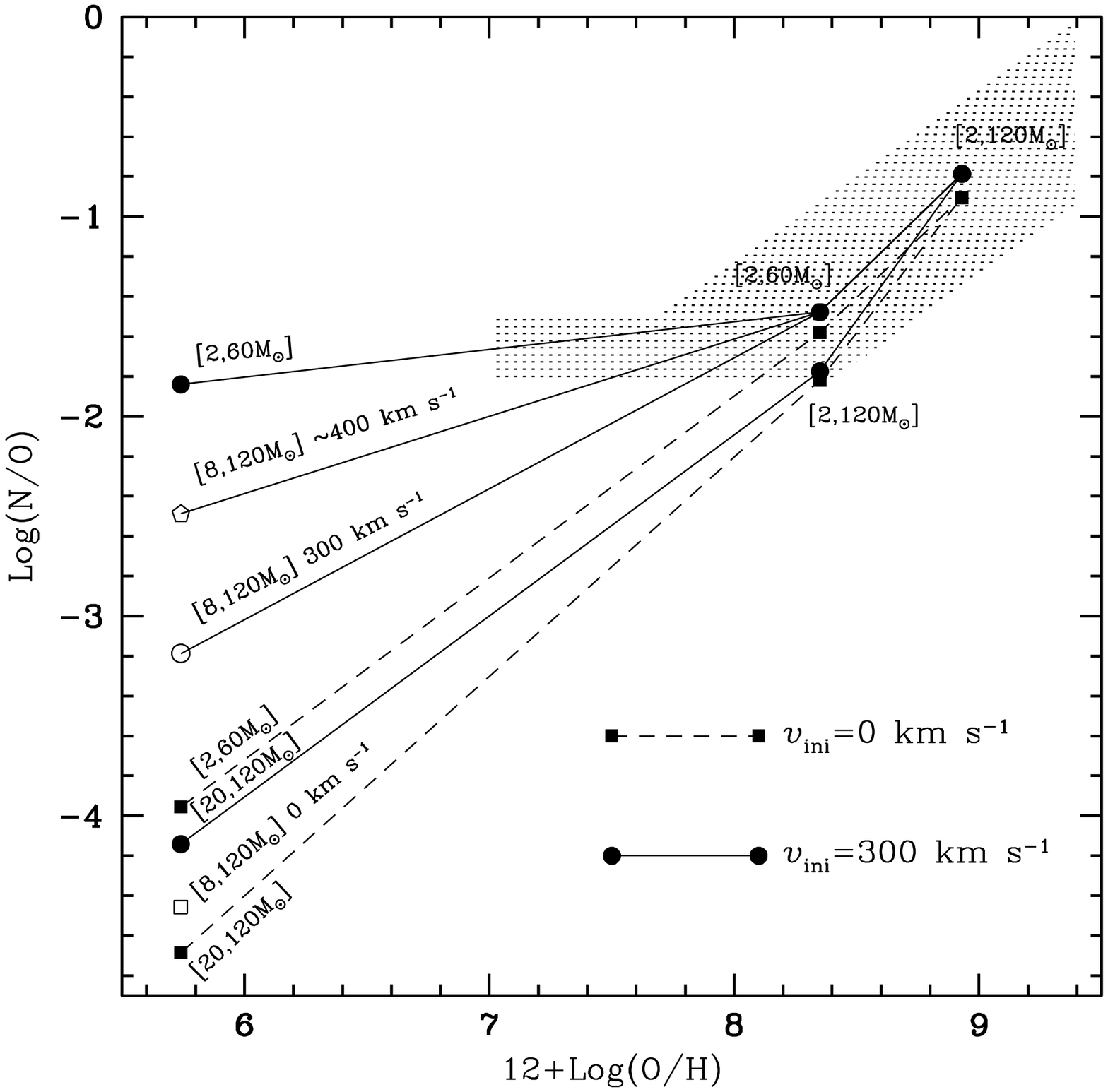}{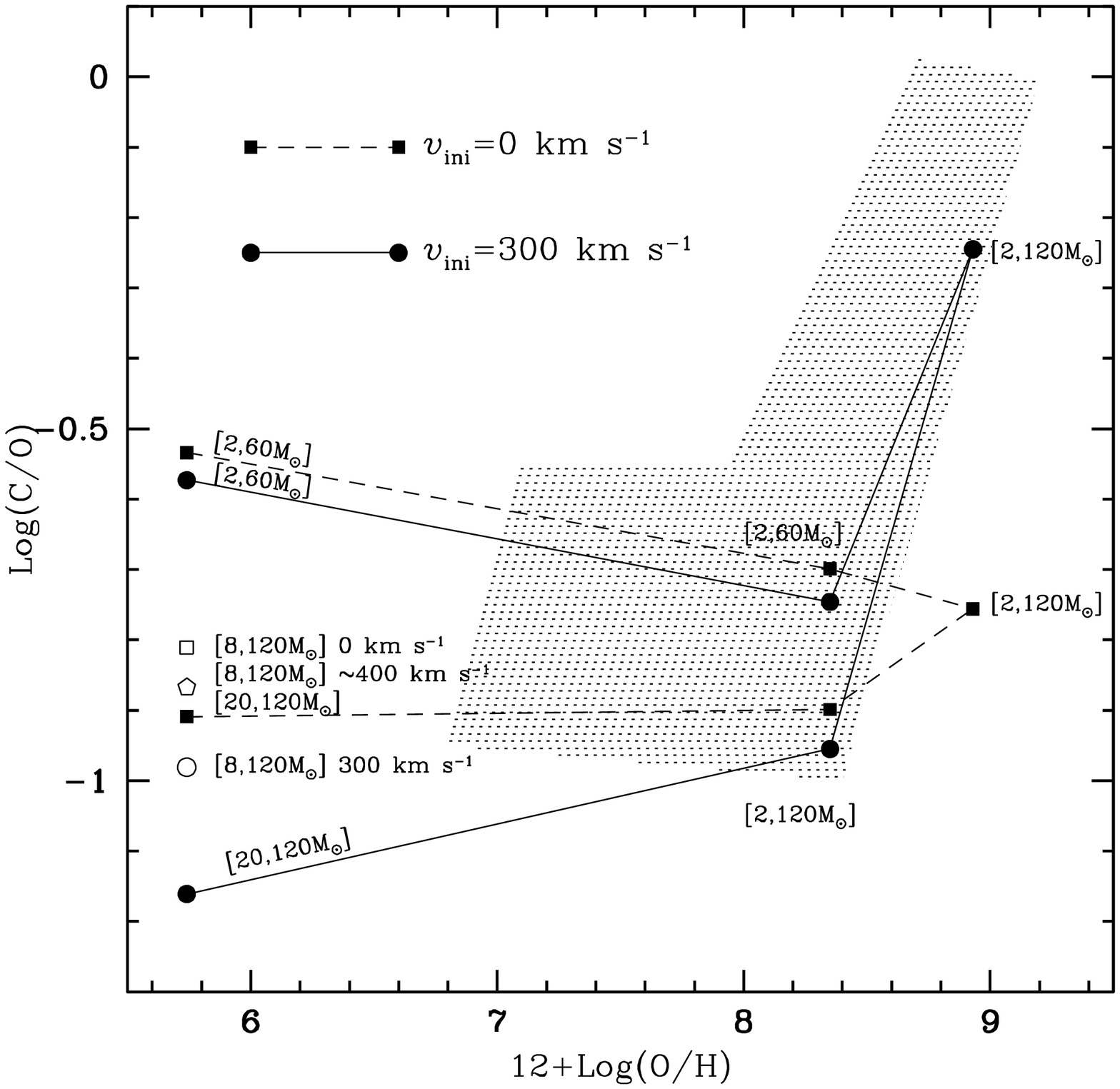}
\caption{{\it Left panel}: Simplified model for the galactic evolution of
the N/O ratio as a function of the O/H ratio (in number). The dashed and 
continuous lines show the results deduced from the non--rotating and the rotating models respectively. 
The range of the initial masses used for computing the integrated yields are indicated.
The empty symbols show the results when only stars more massive than 8 M$_\odot$ are considered.
The initial velocity is indicated.
The shaded area shows the region where most 
of the observations of extragalactic HII regions and stars
are located (see e.g. Gustafsson et al 1999; Henry et al. 2000). {\it Right panel}: Same as left panel for the C/O ratio. The lifetimes
corresponding to these injections are shown in Fig. 7.}
\end{figure}

If, at low metallicity, observations clearly require some source of primary nitrogen, the identification of these sources 
(massive or intermediate mass stars) remains
controversial. Interestingly, there are some observational features which may give some hints: let us suppose that we observe starbursts in galaxies. 
If primary nitrogen is produced by massive stars, one expects to observe a very small scatter of the N/O ratios in starbursts of
different ages, since both nitrogen and oxygen are released at the same time
\footnote{However the observation of a small scatter of the N/O ratios at low metallicity does not necessarily imply
that primary nitrogen is produced by massive stars. Indeed the systems may be sufficiently old
for having allowed intermediate mass stars
to have released their primary nitrogen. In that case, low star formation must be invoked in order to explain the observed low metallicity.}. In case primary nitrogen is released by intermediate mass stars, there will be some time delay between the oxygen and the nitrogen release. Thus there is some chance to observe systems which have already been enriched in oxygen by the massive stars but which are in the process of being enriched in primary nitrogen by intermediate mass stars. One expects in that case some scatter in the observed values of the N/O ratios. Recently, numerous Damped Lyman Alpha (DLA) systems have been observed with values of the N/O ratios well below the plateau level. This strongly suggests that intermediate mass stars are
responsible for the primary nitrogen production.

What are the predictions of the theoretical stellar models ?
In the classical scenario primary nitrogen is produced during the thermal pulse phase
of AGB stars undergoing HBB (see van den Hoek and Groenewegen 1997; Marigo 2001). The quantities of primary nitrogen produced and expelled depends on various parameters as the mass loss along the AGB and the strength of the third dredge--up. Recently, we have proposed a new scenario involving
rotational diffusion: carbon and oxygen produced in the He--core migrate by rotational diffusion in the H--burning shell where they are
transformed into primary nitrogen. In contrast to the classical scenario, this process occurs in the whole mass range and 
is thus not restricted to the intermediate mass range.

The left panel of Fig. 5 shows the chemical abundances as a function of the lagrangian mass in a rotating 3 M$_\odot$ star model at $Z=0.00001$ at the beginning
of the Thermal Pulse--AGB phase. The most striking feature is the very high $^{14}$N abundance obtained at the border of the He--core and in the 
outer convective zone. Its abundance is more than 70 times the initial metal content of the star ! The right panel shows the $^{14}$N stellar yields 
normalised to the initial metal content obtained
for different initial mass stars at various metallicities. 
At very low metallicity, all rotating models produce quantities of new $^{14}$N which are multiples of
the initial metal content. 
 However, for the initial velocities considered here ($v_{\rm ini}$ = 300 km s$^{-1}$), the intermediate mass stars are the main producers.
For higher metallicities, 
the secondary channel for
nitrogen production is the dominant one.

Why the process of primary nitrogen production only works at low metallicity ? 
At low metallicity rotational mixing of chemical species is more efficient due to the steeper gradients of angular velocity inside the stars. These steeper gradients arise as a result of less efficient transport by meridional circulation of the angular momentum in the outer shells
(Meynet \& Maeder 2002ab). In metal poor stars, the H--burning shell is also nearer from the He--burning core. This shortens the timescale
for diffusion between these two zones.

Left panel of Fig. 6 shows the evolution of the N/O ratio as a function of the O/H ratio in the framework of a very simple model for the chemical evolution of galaxies
(closed box model with instantaneous recycling). The hatched zones in Fig. 6 represent regions where many observed points are found (see caption). 
At low metallicity, only the rotating models involving the contribution of the intermediate mass stars
can reproduce the plateau level (at about Log (N/O) = -1.7). Massive stars alone may contribute significantly only if they rotate much faster than
the typical rotational velocities found at solar metallicity. While the evolution in the N/O versus O/H plane mainly depends on rotation, the evolution
in the C/O versus O/H plane is mainly sensitive to the mass range of the stars contributing to the enrichments of the ISM.
Thus, {\it the combination of the diagrams C/O vs O/H, more sensitive to the mass interval, and of the diagram N/O vs. O/H, more
sensitive to rotation, may be particularly powerful to disentangle the two effects of rotation and mass interval}, and to specify
the properties of the star populations responsible for the early chemical evolution of galaxies.

The DLAs present also new and interesting constraints on the source of primary nitrogen at low metallicity.
Recent discussions of this point may be found in the paper of Pettini et al. (2002) as well as in the contributions
by Henry, Molaro in the present volume.
As explained above, the numerous DLAs observed with N/O ratios below the plateau level
favour a primary nitrogen production by intermediate mass stars. Secondly the time delay
required to explain the frequency of the systems with low N/O ratios points towards a relative long time delay between the 
release of oxygen and that of nitrogen (of about 700 Myr according to Pettini et al. 2002). This is well in agreement
with the predictions of the rotating models (see Fig. 7). 

\begin{figure}[t]  
\plottwo{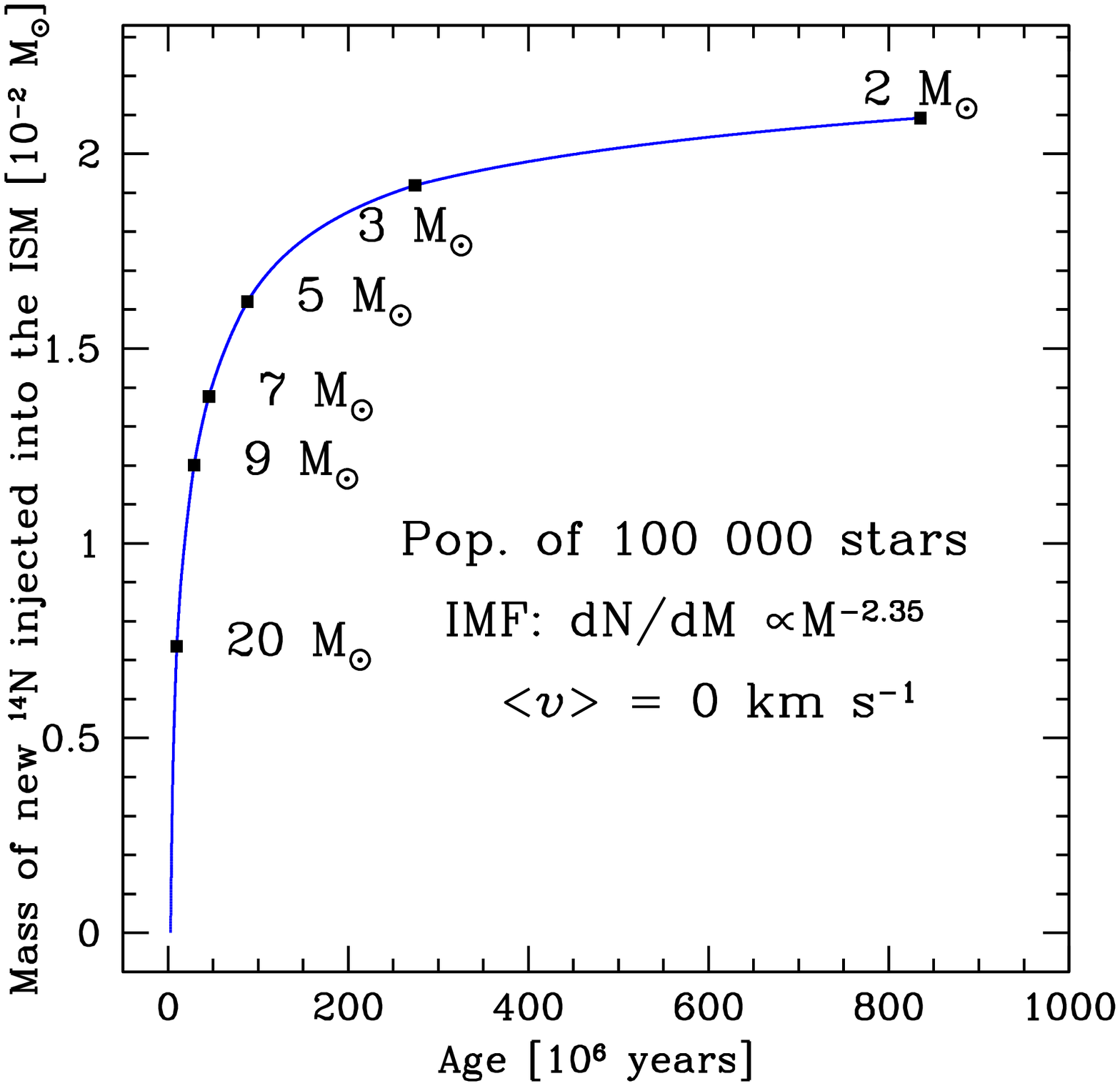}{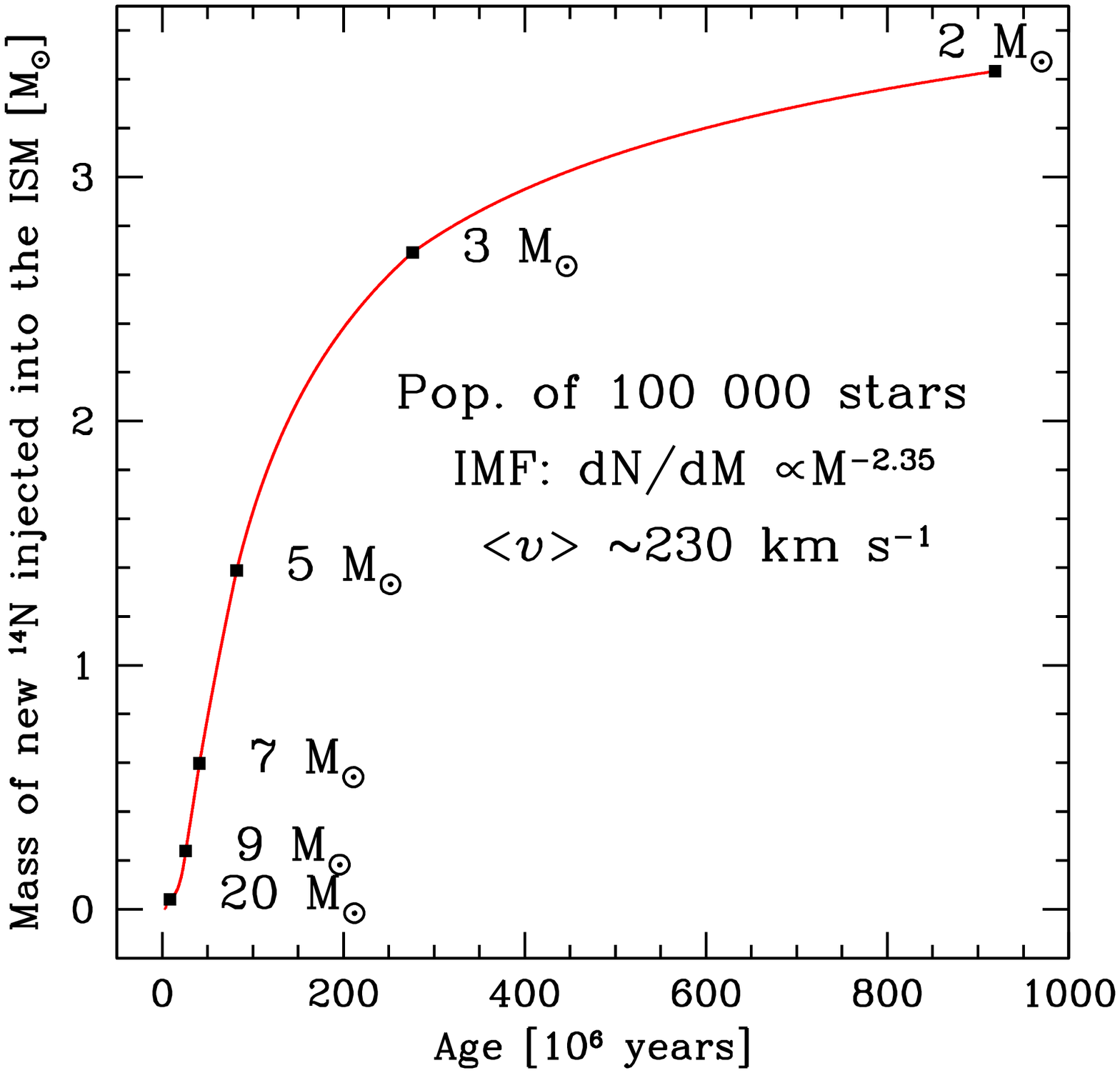}
\caption{Quantities of newly synthesized nitrogen (in solar masses) injected into the interstellar medium by
100 000 stars born at time $t=0$ with initial masses between
0.1 and 60 M$_\odot$. The stellar models have an initial mass fraction of heavy elements equal to 0.00001
(1/2000 the solar metallicity). The left panel shows the situation for non--rotating stellar models, while the right panel
shows the case for rotating stellar models. Note that the vertical axis have different scales in the left and right panel. 
No hot bottom burning process is accounted for. The labels along the curves indicate when a star of a given initial mass contributes.}
\end{figure}

Finally some authors suggest that the N/O values of DLAs are distributed
mainly along two plateaux. One is at Log(N/O) $\sim$ -1.7, as shown on the left panel of Fig. 6. A second plateau
seems to be present
at about 0.7 dex below the first one
(see the papers by Henry and Molaro in this volume). A possible interpretation of the lower plateau is that it 
is produced
by stars with initial masses above 8 M$_\odot$ (see Molaro in this volume). Interestingly, 
the present massive fast rotating  stellar models would be in agreement with this viewpoint. 
As shown on the left panel of Fig. 6, the contribution of rotating stars with masses superior to 8 M$_\odot$ 
and $v_{\rm ini} = 400$ km s$^{-1}$ predict a N/O value at
$Z$ = 10$^{-5}$ about 0.7 dex below the high plateau value.  

\section{Conclusion}

We focused our discussion here on the main isotopes of carbon, oxygen and nitrogen. However rotation may also affect the other less abundant CNO isotopes.
In particular, $^{13}$C and to a lesser extent $^{17}$O may also be abundantly produced in rotating models 
and have a primary production channel at low metallicity. 
As an example, at $Z$ = 0.00001, the amounts of $^{13}$C  synthesized in rotating intermediate mass star models may be nearly 10 000 times the amount
produced in non--rotating models.
Interestingly some primary $^{13}$C production seems to be required by the observations (Prantzos et al. 1996). Important productions
of primary $^{22}$Ne are also predicted by the rotating models.

\end{document}